\documentclass[twocolumn]{aastex631}

\usepackage{commath}
\usepackage{floatrow}
\usepackage{soul}
\usepackage{CJK}
\usepackage{threeparttable}
\usepackage{makecell}
\usepackage{multirow}
\usepackage{txfonts}
\usepackage{CJKulem}

\newcommand{\HII}{\hbox{H\,{\scriptsize II}}}

\received{}
\revised{}
\accepted{}

\submitjournal{APJL}

\shorttitle{$\tau_{\rm{cl}}$ and DTM}
\shortauthors{Lu et al.}
\graphicspath{{./}{figures/}}

\begin{document}
\begin{CJK*}{UTF8}{gbsn}

\title{The\textit{ Chocolate Chip Cookie} Model: dust-to-metal ratio of $\HII$ regions }

\author[0000-0002-8817-4587]{Jiafeng Lu (卢家风)}
\affiliation{Key Laboratory for Research in Galaxies and Cosmology, Shanghai Astronomical Observatory, Chinese Academy of Sciences, 80 Nandan Road, Shanghai 200030, People's Republic of China}
\affiliation{University of Chinese Academy of Sciences, 19A Yuquan Road, Beijing 100049, People's Republic of China}

\author[0000-0002-3073-5871]{Shiyin Shen (沈世银)}
\affiliation{Key Laboratory for Research in Galaxies and Cosmology, Shanghai Astronomical Observatory, Chinese Academy of Sciences, 80 Nandan Road, Shanghai 200030, People's Republic of China}
\affiliation{Key Lab for Astrophysics, Shanghai, 200034, People's Republic of China}

\author[0000-0001-6763-5869]{Fang-Ting Yuan (袁方婷)}
\affiliation{Key Laboratory for Research in Galaxies and Cosmology, Shanghai Astronomical Observatory, Chinese Academy of Sciences, 80 Nandan Road, Shanghai 200030, People's Republic of China}
\affiliation{Key Lab for Astrophysics, Shanghai, 200034, People's Republic of China}

\author[0000-0002-4638-2580]{Qi Zeng (曾琪)}
\affiliation{Key Laboratory for Research in Galaxies and Cosmology, Shanghai Astronomical Observatory, Chinese Academy of Sciences, 80 Nandan Road, Shanghai 200030, People's Republic of China}
\affiliation{University of Chinese Academy of Sciences, 19A Yuquan Road, Beijing 100049, People's Republic of China}

\begin{abstract}
 Using a sample of face-on star-forming galaxies selected from the Sloan Digital Sky Survey, we statistically derive the typical optical depth $\tau_{\rm{cl}}$ of individual $\HII$ regions based on the ``Chocolate Chip Cookie" model of \citet{Lu2022}. By binning galaxies into stellar mass and gas-phase metallicity bins and  interpreting $\tau_{\rm{cl}}$ as the dust to gas ratio (DGR) of $\HII$ regions, we further investigate the correlations among DGR and stellar mass, gas-phase metallicity respectively. We find that DGR increases monotonically with the stellar mass of galaxies. At a given stellar mass,  DGR shows a linear correlation with the gas-phase metallicity, which implies a constant dust to metal ratio (DTM) of galaxies at a given stellar mass. These results adequately indicate that the DTM of galaxies is simply a function of their stellar masses. In terms of gas-phase metallicity, because of the mass-metalliciy relation, DTM increases with increasing metallicity with a power-law index 1.45 in the low metallicity region, while remains constant at the high metallicity end. 
\end{abstract}


\section{Introduction} \label{sec:intro}

Dust plays an important role in star formation and galaxy evolution. The properties of dust in galaxies, particularly the dust-to-gas ratio (DGR) and dust-to-metal ratio (DTM) have been intensively studied through statistical scaling relations \citep{Lisenfeld1998,Hirashita2002,Draine2007,Galametz2011,Zafar2013,Remy2014,De_Vis2019,Wiseman2017,Kahre2018}. Dust evolution models show that the growth of dust grains increases  DTM \citep{Mattsson2012} , while dust destruction performs in the opposite way \citep{Draine1979}.  Therefore, in observation, the DGR or DTM as a function of metallicity  provides an effective constraint on dust evolution mechanisms.  Early studies have suggested a linear relation between DGR and metallicity, or a constant DTM \citep{Issa1990,Lisenfeld1998}, for a wide range of galaxies, which was explained by the ineffectiveness of grain growth and destruction \citep{Hirashita1999,Edmunds2001}. A constant DTM is also assumed in many semi-analytic models and hydro-dynamical simulations \citep{Silva1998,James2002,Clark2016,Yajima2015,Camps2015,Somerville2012,Ma2019,Katz2019}. In recent studies, a variable metallicity dependent DTM has been widely presented by infrared (IR) observations, especially for galaxies with low metallicity \citep[for example][]{Remy2014,De_Vis2017,De_Vis2019}. However,  selection effects or uncertainties could result in different observational trends because of the small sample size of a few hundred galaxies \citep{Mattsson2014}.
 
The large galaxy samples in optical survey is more suitable for statistical study. However, it is not straightforward to derive the DGR or DTM from the optical observations alone. With spectroscopic observations, the gas-phase metallicity of star forming galaxies (SFG) can be easily derived from the emission lines of $\HII$ regions \citep[for example][]{Tremonti2004}. Therefore, if we can have a implicit  probe of the average amount of dust (i.e. optical depth) of the $\HII$ regions of SFG using optical observations, we could study the dependence of DGR (or DTM) on the other physical properties of galaxies statistically with a large sample. However, the statistical estimation of the average amount of dust (optical depth) of  $\HII$ regions in extra-galactic galaxies is nontrivial. First of all, we can not get the observational constraint on the optical depth of $\HII$ regions through the reddening of the emission lines directly.  The emission lines of $\HII$ regions are not only  extincted by the dust layer of themselves, but also extincted by the foreground ISM along the line-of-sight.  Moreover, considering the clumpy distribution of $\HII$ regions, along the line-of-sight, especially when the disk galaxies are highly-inclined,  there will be a inter-covering effect of $\HII$ regions, which will certainly result in a biased estimation of the average dust extinction of $\HII$ regions. 

Recently, \citet{Lu2022} have provided a framework of the dust configuration of disk galaxies named as the  ``Chocolate Chip Cookie" (hereafter CCC) model, in which the clumpy nebular regions are embedded in a diffuse stellar/ISM disk, like chocolate chips in cookies.  In this model, the average optical depth of individual $\HII$ regions $\tau_{\rm{cl}}$ is a model parameter, which can be well constrained because both the foreground ISM dust extinction and the mutual attenuation of $\HII$ clumps have been carefully considered. Moreover, as we will show in Section \ref{sec:sample}, for face-on galaxies, the derivation of the optical depth of individual $\HII$ regions could be even simplified. That is to say, using the CCC model, we can obtain the typical optical depth of the $\HII$ regions and then estimate the DTM for a given sample of star-forming galaxies based on the optical observation data alone. 

In this work, we aim to derive the average optical depth of individual $\HII$ regions, $\tau_{\rm{cl}}$, via an approximation of CCC model with a sample of face-on SFG in the Sloan Digital Digital Sky Survey (SDSS). Then, we explore the dependence of $\tau_{\rm{cl}}$ on the stellar mass and gas-phase metallicity of galaxies, which gives us more details about the variations of DTM and puts constraints on the possible physical mechanisms of these variations. 

The outline of this paper is as follows. In Section \ref{sec:sample}, we present a face-on SFG sample and show their dust reddening features. In Section \ref{sec:method}, we introduce a method to interpret DTM as a function of $\tau_{\rm{cl}}$ and gas-phase metallicity. Then, we use the simplified ``CCC model" to derive  $\tau_{\rm{cl}}$ for different sub-sample of galaxies.  We present our main results in Section \ref{sec:result} and make discussions on the DTM of SFGs in Section \ref{sec:discussion}. Finally, we make a summary in Section \ref{sec:conclusion}.

\section{Data: face-on disk galaxy sample} 
\label{sec:sample}

In this study, we use the SFG sample of \citep{Lu2022}, which was selected from the spectroscopic main sample galaxies of the SDSS using the BPT diagram applied with the criteria of \citet{Kauffmann2003a} and requiring the signal-to-noise ratio (S/N) of each emission lines larger than 3.  Unlike the Milky-Way like (MW-like) galaxies studied in \citet{Lu2022}, we apply the CCC model to galaxies in a much large stellar mass range ($9< \log \rm{M_*} <11$, where $\rm{M_*}$ represents the stellar mass in unit of solar mass). Moreover, we further select the face-on SFGs with an axis ratio $b/a > 0.8$ to simplify the CCC model so that the  constraints on the optical depth of the $\HII$ regions can be more explicitly obtained (see Section \ref{sec:method} for detail). 
We therefore finally obtain a sample of 25,573 face-on SFGs within the stellar mass range $9<\log \rm{M_*} <11$. In our sample, the gas-phase metallicity (12+$\log (\rm{O/H})$) is obtained from the MPA-JHU database\footnote{https://wwwmpa.mpa-garching.mpg.de/SDSS/DR7/oh.html}, following \citet{Tremonti2004}.

Following the same method as \citet{Lu2022}, we first obtain the stellar reddening $E_{\rm{s}}$ and the emission line reddening $E_{\rm{g}}$ for each sample galaxies. In short, the stellar reddening $E_{\rm{s}}$ is obtained via  the stellar population synthesis code STARLIGHT \citep{Fernandes2005} with the BC03 single stellar population \citep{Bruzual2003}. The emission line reddening $E_{\rm{g}}$ is obtained via Balmer decrement. 

\begin{figure}[htbp]
\centering
\gridline{\fig{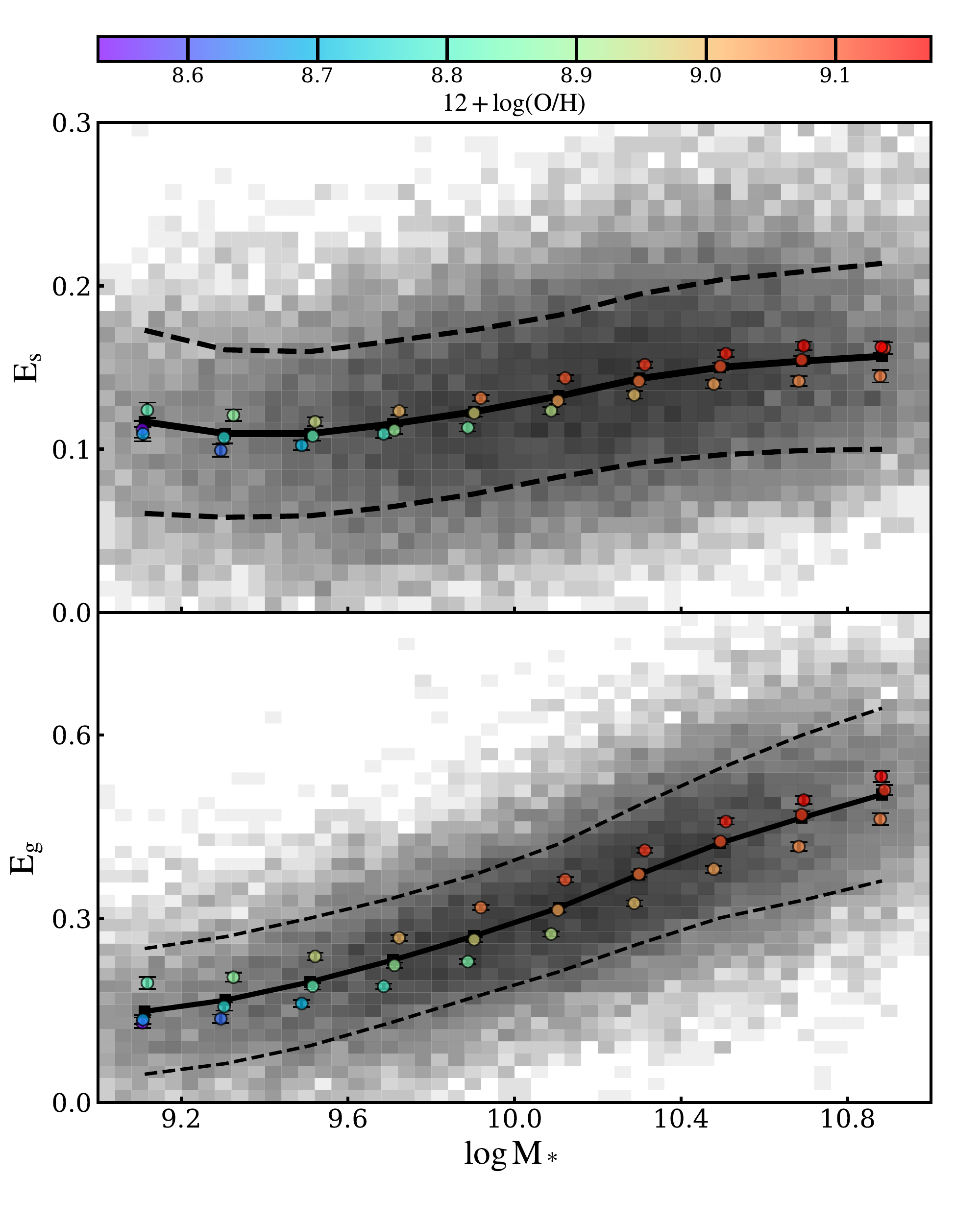}{3.5in}{}
          }
      
 \caption{The  stellar reddening $E_{\rm{s}}$ (top panel) and nebular reddening $E_{\rm{g}}$ (bottom panel) as a function of stellar mass for  face-on SFGs. In each panel, the black line  represents the median $E_{\rm{s}}$ (or $E_{\rm{g}}$), where the two black  dashed lines are the 16 and 84 percentiles of $E_{\rm{s}}$ (or $E_{\rm{g}}$)  respectively. For each stellar mass bin, the colored dots further show the  median $E_{\rm{s}}$ (or $E_{\rm{g}}$) of three different metallicity bins, which are color-coded  according to the color-bar plotted on the top of the figure. }
\label{fig1} 
\end{figure}

We show the resulted $\rm{M_*}\mbox{-}E_{\rm{s}}$ and $\rm{M_*}\mbox{-}E_{\rm{g}}$ density distribution in Figure \ref{fig1}. The medians of $E_{\rm{g}}$ and $E_{\rm{s}}$ in stellar mass bins of 0.2 dex are connected  with black solid lines, while the black dashed lines are the contours that enclose 68 percentiles of $E_{\rm{g}}$ or $E_{\rm{s}}$ distribution.  Moreover, at a given stellar mass bin, we further divide the sample galaxies into 3 metallicity bins with equal number of galaxies and calculate their median $E_{\rm{s}}$ and $E_{\rm{g}}$, respectively, which are shown as the color dots in Figure \ref{fig1}. We also estimate the uncertainties of the median $E_{\rm{g}}$ and $E_{\rm{s}}$ in each stellar mass and metallicity bin using the formula $\sqrt{\frac{\pi}{2}}\frac{\sigma}{\sqrt{n}}$, where $\sigma$ is the standard deviation of $E_{\rm{s}}$ or $E_{\rm{g}}$ distribution and $n$ is the number of galaxy sample in each bin. These uncertainties are also plotted in Figure \ref{fig1} as the errorbar on each color dot and will be used to estimate the uncertainties $\tau_{\rm{cl}}$  in Section \ref{sec:method}. As shown in  Figure \ref{fig1},  $E_{\rm{s}}$ shows a plateau at 0.1 for low stellar mass ($\log\rm{M_*}<9.5$), then increases slightly from 0.1 to 0.15 with stellar mass in the range of $9.5<\log \rm{M_*}<11$, while  $E_{\rm{g}}$ increases monotonically from 0.1 to 0.5 with stellar mass. At given stellar mass, both $E_{\rm{g}}$ and $E_{\rm{s}}$ further increase with the increasing of metallicity.

It is important to note that our measurement of  $E_{\rm{g}}$ does not directly represent the dust extinction (optical depth) of the $\HII$ region itself. However, with these two measurements ($E_{\rm{g}}$ and $E_{\rm{s}}$), as we will show in Section \ref{sec:method}, our CCC model can easily and robustly estimate $\tau_{\rm{cl}}$ for face-on SFGs.

\begin{figure}[htbp]
\centering
\gridline{\fig{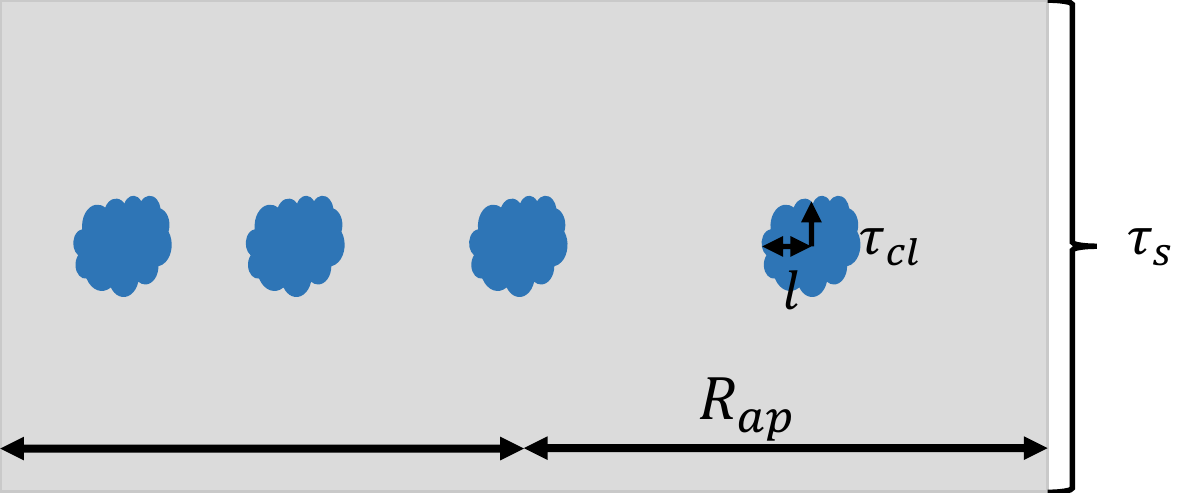}{3.5in}{}
          }
 \caption{Sketch diagram of simplified CCC model for a face-on disk galaxy inside a fiber aperture (radius $R_{\rm{ap}}$), where a single layer of individual $\HII$ regions (blue clouds, intrinsic optical depth $\tau_{\rm{cl}}$ and dust shell thickness $l$) are sparsely embedded in the middle of a thick  ISM  disk (gray background, total optical depth $\tau_{\rm{s}}$ along the line-of-sight). }
\label{fig2} 
\end{figure}

\section{Method: deriving optical depth of individual \texorpdfstring{$\HII$}{HII} regions }
\label{sec:method}

Before deriving the typical optical depth $\tau_{\rm{cl}}$ of individual $\HII$ regions, we first discuss how  $\tau_{\rm{cl}}$ could represent the dust abundance of $\HII$ regions. We assume a simple model of a sphere with a dust shell for $\HII$ regions. For the inner sphere, the gas is fully ionized, and we assume there is no dust. In the outer shell, the dust is uniformly mixed with the neutral gas. The metallicity is assumed to be the same for the ionized gas of the inner sphere and the neutral gas of the outer shell. With this toy model, We write $\tau_{\rm{cl}}$ in the form

\begin{equation}
    \tau_{\rm{cl}}=\kappa_{V} \rho_{\rm{dust}} l ,
     \label{tcl1}
\end{equation}

where  $\rho_{\rm{dust}}$ is the volume density of the dust particles in the outer shell, $l$ is the thickness of the dust shell, $\kappa_V$ is the dust absorption coefficient.
By linking $\rho_{\rm{dust}}$ to the DTG and neutral gas volume density of shell $\rho_{\rm{gas}}$, we get 
\begin{equation}
    \tau_{\rm{cl}}=\kappa_{V} \rho_{\rm{gas}} l \cdot \mathrm{DTG}\,.
    \label {tcl2}
\end{equation}
In equation \ref{tcl2}, $\rho_{\rm{gas}}$ and $l$ parameterize the basic geometric properties of shell of $\HII$ regions, $\kappa_V$ characterizes the physical properties of dust particles.  

Here, we assume that $\rho_{\rm{gas}}$, $l$ and $\kappa_V$ are all independent of their host galaxies. Considering the the average electron density and size distribution of $\HII$ regions are similar for all types of local SFGs \citep{Oey2003, Liu2013, Santoro2022} and the inner ionized sphere and dust shell are both a part of $\HII$ complex, the assumption is somewhat reasonable. We therefore get $\tau_{\rm{cl}}\propto$ DTG. With this assumption, we will use $\tau_{\rm{cl}}$ to represent the DTG of galaxies and then study its dependence on other physical properties of galaxies.

The DTG can be further written as the product of DTM (denoted as $\zeta$) and metallicity $Z$. In this study,  we use the gas-phase metallicity estimated by  \citet{Tremonti2004}, which is defined as the logarithmic abundance ratio of oxygen to hydrogen and denoted as $12+\log( \rm{O/H})$. Therefore, in logarithmic space, the relation between $\zeta$ and $\tau_{\rm{cl}}$ is 
\begin{equation}
C + \log \zeta =\log \tau_{\rm{cl}} - \log( \rm{O/H}). \,
\label {tcl3}
\end{equation}
where the constant term $C$ represents the unknown and unrelated constant factors assumed in this study (e.g. $\kappa_V, \rho_{\rm{gas}},l$ in Equation \ref{tcl2}).

In \citet{Lu2022}, we have presented a  two-component dust geometry model for disk galaxies, the CCC model, where the $\HII$ regions (``chocolate chips") are embedded in a continuously distributed ISM disk (``cookie"). In the CCC model, the diffuse ISM is assumed to be uniformly mixed with the stellar component, whereas the $\HII$ regions are clumpy and distributed in a much thinner disk.

For face-on galaxies, this model can be further simplified to a scenario in which  a single layer of $\HII$ regions sparsely embedded in the middle of a thick ISM disk, as illustrated in Figure \ref{fig2}. In \citet{Lu2022}, we have obtained the cross section of $\HII$ regions in the central region of MW-like galaxies is 0.84 $\rm{kpc}^{-1}$ (defined as the average number of $\HII$ regions in the central line-of-sight) and a very low thickness of the $\HII$ disk (scale-height $\sim 0.1$ kpc). 
For face-on MW-like galaxies, the covering factor of the $\HII$ regions inside a fiber aperture is about 0.1 so that the obscuration of the $\HII$ regions to the stars is negligible. In this study, we assume all SFGs have a dust geometry configuration similar to MW-like galaxies and make two reasonable assumptions for all face-on SFGs: (1) there is at most one $\HII$ region along any single line-of-sight; (2) the covering factor of $\HII$ regions is very low. 

With these two assumptions, the stellar emission is only attenuated by the ISM dust and the $\HII$ nebular emission is attenuated by the ISM dust and the dust shell of the $\HII$ region itself. Since the ISM dust is assumed to be uniformly mixed with stellar population in the CCC model, the stellar attenuation and the emission line attenuation can be simplified as the uniform mixture model and the screen model respectively, which have also been discussed in \citet{Lu2022}. Besides, as the typical size of $\HII$ regions is much smaller than the scale-height of the stellar disk, the optical depth of the foreground ISM dust of $\HII$ regions can be approximated by the half of the total optical depth of ISM dust. Finally, the stellar and nebular reddening are then both the functions of optical depths:
\begin{equation}
\begin{aligned}
&E_{\rm{s}} =2.5\log(\frac{\tau_{\rm{s},B}}{\tau_{\rm{s},V}}\frac{1-e^{-\tau_{\rm{s},V}}}{1-e^{-\tau_{\rm{s},B}}}) 
\\
&E_{\rm{g}}=1.086(\tau_{\rm{g},\rm{H}\beta}-\tau_{\rm{g},\rm{H}\alpha})
\\
& \tau_{\rm{g}}=\frac{\tau_{\rm{s}}}{2}+\tau_{\rm{cl}} \,.
\end{aligned}
\label{eq4}
\end{equation}
where $\tau_{\rm{s}}$ is the total optical depth of ISM dust(see Figure \ref{fig2}), $\tau_{\rm{g}}$ is the total optical depth of the nebular region along the line-of-sight and $\tau_{\rm{cl}}$ is the optical depth of individual $\HII$ regions only.

Based on Equation \ref{eq4},  $\tau_{\rm{s}}$ and $\tau_{\rm{cl}}$ \footnote{In this paper, if not explicitly specified, the optical depth is defined in the default wavelength $V$-band.} can be derived from the observed $E_{\rm{g}}$ and $E_{\rm{s}}$ by assuming an extinction curve. Following \citet{Lu2022}, we assume a power law attenuation curve with the typical total to selective extinction $R_V=3.1$ \citep{Li2017,Fitzpatrick2019}:
\begin{equation}
\frac{A_{\lambda}}{A_{V}}=(\frac{\lambda}{5500\text{\AA}})^{-1.32} 
\label{eq:power curve}
\end{equation}
As we have shown in \citet{Lu2022},  when given a certain geometry of dust and stars, this extinction curve can reproduce the classical attenuation curve of Calzetti law \citep{Calzetti2000}. 

It should be noted that the CCC model deduces $\tau_{\rm{cl}}$ of individual $\HII$ regions from a statistical perspective. Moreover, instead of deriving $\tau_{\rm{cl}}$ for each individual galaxy using Equation \ref{eq4}, we take the median $E_{\rm{g}}$ and $E_{\rm{s}}$ for a sample of galaxies with similar physical properties and calculate their typical $\tau_{\rm{cl}}$ values. Meanwhile, the uncertainties of $\tau_{\rm{cl}}$  are derived by propagating the uncertainties of $E_{\rm{g}}$ and $E_{\rm{s}}$. In other words, the motivation of this paper is to study the $\tau_{\rm{cl}}$ of SFGs in a statistical sense, rather than focusing on the specific values of $\tau_{\rm{cl}}$ for any individual galaxies.

\section{Results}

\label{sec:result}

\begin{figure*}[htbp]
\centering
\gridline{\fig{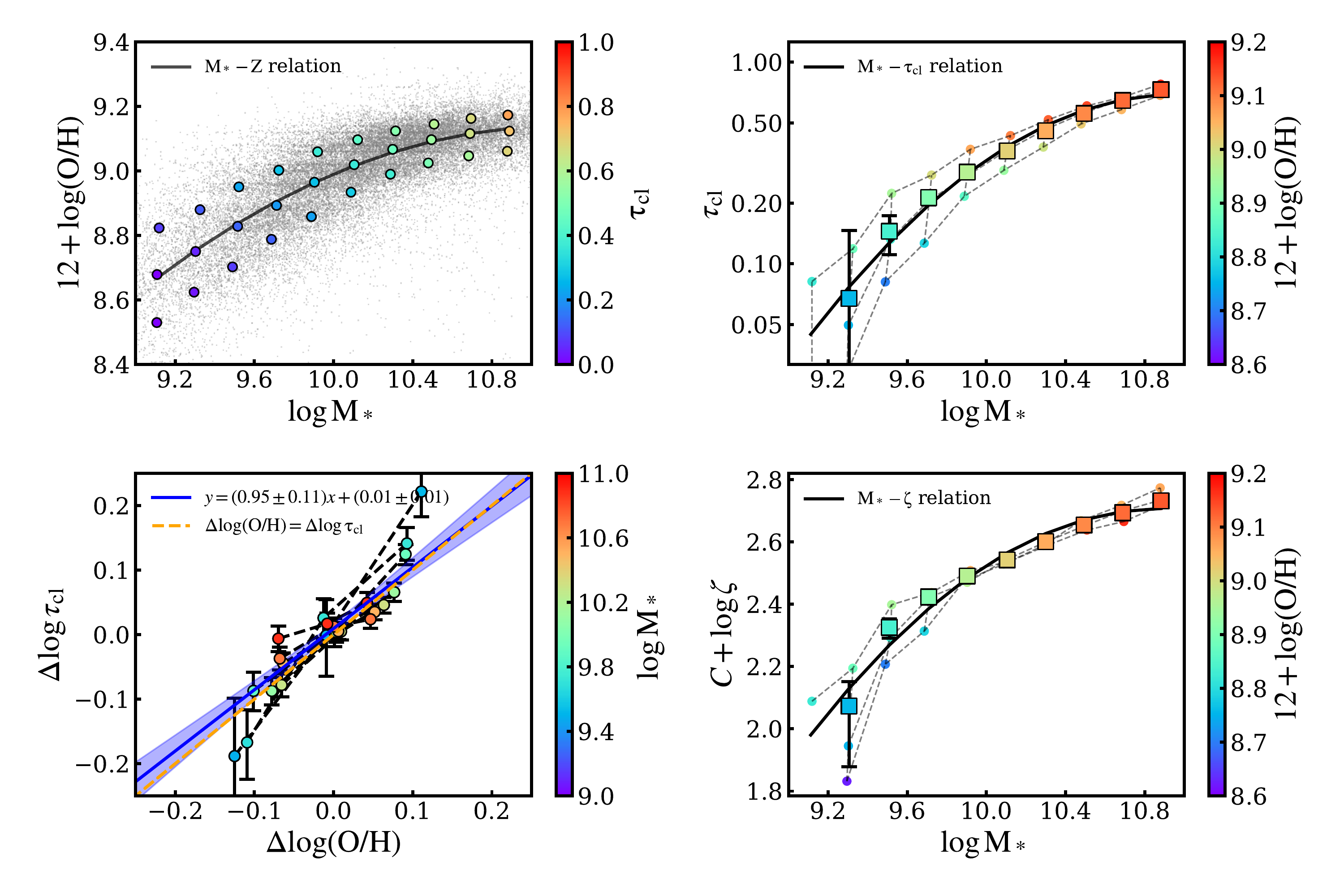}{7in}{}
          }
 \caption{The dependence of $\tau_{\rm{cl}}$ on stellar mass $\rm{M_*}$ and metallicity 12+$\log (\rm{O/H})$. Top-left panel:  $\rm{M_*}\mbox{-}Z$ relation,  and $\tau_{\rm{cl}}$ is color-coded. The circles represent the 30 sub-sample of galaxies in $\rm{M_*}$ and $Z$ bins, while the small black dots in background are all sample galaxies. The solid curve is the the fitted $\rm{M_*}\mbox{-}Z$ relation of Equation \ref{eq6}. Top-right panel: $\rm{M_*}\mbox{-}\tau_{\rm{cl}}$ relation, and metallicity color-coded. The dots connected by dotted lines represent the sub-sample of galaxies in 30 stellar mass and metallicity bins, while the squares show the median $\tau_{\rm{cl}}$ values in 10 stellar mass bins irrespective of metallicity. The solid curve is the fitted $\rm{M_*}\mbox{-}\tau_{\rm{cl}}$ relation of Equation \ref{eq7}. Bottom-left panel: the residual $\Delta \log \tau_{\rm{cl}} \mbox{-} \Delta \log(\rm{O/H})$ relation color-coded with the $\rm{M_*}$ of different stellar mass bin. The galaxy bins with the same stellar mass are connected by black dashed lines. The best fit and the confidence level of 1$\sigma$ are shown by the blue solid curve and shadowed regions. The dashed orange diagonal line ($\Delta \log \tau_{\rm{cl}} = \Delta \log(\rm{O/H})$) is plotted for comparison. Bottom-right panel: $\rm{M_*}\mbox{-}\zeta$ relation color-coded with the metallicity.  The data points are  the same as that of the upper left panel. The solid curve follows Equation \ref{DTMM}.}
\label{fig3} 
\end{figure*}

In Section \ref{sec:sample}, we divide the face-on SFGs into 10 stellar mass bins with bin width of 0.2 dex, and then the galaxies at given stellar mass bin are further divided into sub-samples of 3 gas-phase metallicity bins with equal numbers. In this section, we take these sub-samples and  explore the variation of $\tau_{\rm{cl}}$ as function of stellar mass $\rm{M_*}$ and metallicity ($12+\log (\rm{O/H})$) of host galaxies using the simplified CCC model outlined in Section \ref{sec:method}. 

We first show the $\rm{M_*}\mbox{-}Z$ relation of all face-on SFGs with small dots in the the top-left panel of Figure \ref{fig3}. We show the median $12+\log(\rm{O/H})$ as a function of stellar mass with the solid curve, which is parameterized by a  quadratic function following \citet{Tremonti2004}\footnote{This fitting formula is obtained using curve$\_$fit module in scipy library of Python3, and is very similar to that of \citet{Tremonti2004}. }:
\begin{equation}
\begin{aligned}
    &12+\overline{\log(\rm{O/H})}(\rm{M_*})=\\
    &-0.114 (\log \rm{M_*})^2+2.534 \log \rm{M_*}-4.978 \,.
\label{eq6}
\end{aligned}
\end{equation}
Besides, the subsamples of galaxies in 30 stellar mass and metallicity bins are shown by circle dots in this panel, whose color codes the $\tau_{cl}$ values of each subsample. As can be seen, more massive and more metal rich galaxies have systematically higher $\tau_{\rm{cl}}$ values, ranging from $\tau_{\rm{cl}}\sim 0.05$ for the least massive and metal poor galaxies ($\log \rm{M_*}\sim 9, 12+\log(\rm{O/H}) \sim 8.5$) to $\tau_{\rm{cl}}\sim 0.8$ for the most massive and metal rich galaxies ($\log \rm{M_*}\sim 11,  12+\log(\rm{O/H}) \sim 9.2$). 

To show the dependence of $\tau_{\rm{cl}}$ on stellar mass and metallicity clearly, we plot $\tau_{\rm{cl}}$ as a function of $\rm{M_*}$ in the top-right panel of Figure \ref{fig3}, where the median $12+\log(\rm{O/H})$ of each sample galaxies are color coded. We first show the median $\tau_{\rm{cl}}$ of sample galaxies in 10 stellar mass bins irrespective of their metallicity as big squares. The $\tau_{\rm{cl}}$ of sample galaxies in 30 stellar mass and metallicity bins are then shown as circle dots connected by dotted lines. For clarity, we only show the uncertainties of $\tau_{\rm{cl}}$ for 10 stellar mass bins. For the global median $\tau_{\rm{cl}}$ in 10 $\rm{M_*}$ bins (squares), we see that $\overline{\tau_{\rm{cl}}}$ increases monotonically with stellar mass. Following the $\rm{M_*}\mbox{-}Z$ relation, we also parameterize the $\rm{M_*}\mbox{-}\overline{\tau_{\rm{cl}}}$ relation with a quadratic function  and get the best fit 
\begin{equation}
    \log \overline{\tau_{\rm{cl}}} (\rm{M_*}) = -0.343(\log \rm{M_*})^2+7.54 \log \rm{M_*}-41.531,
    \label{eq7}
\end{equation}
which is shown as the black solid curve in this panel. Besides the $\rm{M_*}\mbox{-}Z$ degeneracy, again, we see a clear second-order dependence of $\tau_{\rm{cl}}$ on metallicity at given stellar mass: higher $Z$ galaxies have higher $\tau_{\rm{cl}}$.

To show the second order dependence of $\tau_{\rm{cl}}$ on $Z$ more clearly, we plot the residual and their error of the $\rm{M_*}\mbox{-}\tau_{\rm{cl}}$ relation ($\Delta \log \tau_{\rm{cl}}$) as function of the residual of the $\rm{M_*}\mbox{-}Z$ relation ($\Delta \log(\rm{O/H})$) in the bottom-left panel of Figure \ref{fig3}.  In specific, these two residuals are defined as
\begin{equation}
\begin{aligned}
   \Delta \log(\rm{O/H}) &=\log(\rm{O/H})-\overline{\log(\rm{O/H})}(\rm{M_*}), 
\\
   \Delta\log\tau_{\rm{cl}}&=\log\tau_{\rm{cl}}-\log\overline{\tau_{\rm{cl}}}(\rm{M_*})\,
\label{eq8}
\end{aligned}
\end{equation}
where $\overline{\log(\rm{O/H})}(\rm{M_*})$ and $\log \overline{\tau_{\rm{cl}}}(\rm{M_*})$ are defined by Equations \ref{eq6} and \ref{eq7}, respectively. We also fit the global trend with a linear relation. The relation and its confidence level of 1$\sigma$ are also plotted with blue solid line and shadowed regions in the bottom-left panel of Figure \ref{fig3}. As can be seen, after subtracting the $\rm{M_*}\mbox{-}\overline{\tau_{\rm{cl}}}$ relation and $\rm{M_*}\mbox{-}\overline{Z}$ relation, $\Delta \log \tau_{\rm{cl}}$ roughly shows a linear relation with $\Delta \log(\rm{O/H})$ for all $\rm{M_*}$ bins with a slope of 1. According to Equation \ref{tcl3}, this equivalence implies that the DTM ($\zeta$) of galaxies with different metallicity are roughly a constant at given stellar mass.

To further investigate the results shown above, we plot $\zeta$ (defined as Equation \ref{tcl3}) as a function of $\rm{M_*}$ only in the bottom-right panel of Figure \ref{fig3}, color coded according to the metallicity. Following the top-right panel, we show the sample galaxies in 10 stellar mass bins irrespective of their metallicity as big squares and these sub-samples in 30 stellar mass and metallicity bins as circle dots connected by dotted lines. As can be seen, $\zeta$ almost shows no second-order dependence on metallicity for most stellar mas bins ($\log \rm{M_*}>9.8$). For these few low mass bins, considering the uncertainties of $\zeta$ there, we believe $\zeta$ is also largely independent of metallicity. Even if there is a correlation, the correlation should be very weak. 
Since $\zeta$ is only a function of stellar mass, we provide an analytic expression of that function. Rather than fitting the data points in the bottom-right panel, we can also obtain it by the combination of Equations \ref{tcl3}, \ref{eq6}, \ref{eq7} and \ref{eq8}, 

\begin{equation}
\begin{aligned}
    C + \log \zeta &= \log \overline{ \tau_{\rm{cl}}}(\rm{M_*})- \overline{\log(\rm{O/H})}(\rm{M_*})
    \\
     &= -0.229(\log \rm{M_*})^2+5.006 \log \rm{M_*}-24.553 \,.
    \label{DTMM}
\end{aligned}
\end{equation}
We plot this quadratic function as the solid curve in this panel. As can be seen, this derived analytic formula  matches the data points very well.

From the plots shown above, we arrive at the main finding of this study: the DTM of a galaxy is a unique function of the stellar mass of its host galaxy. This finding has important implications for the time-scale of dust particle evolution, which will be further discussed in Section \ref{sec:zeta-M}.

\section{Discussion}
\label{sec:discussion}

Most of the studies on dust evolution model focus on the dependence of $\zeta$ on metallicity \citep{Lisenfeld1998,Hirashita2002,Draine2007,Galametz2011,Zafar2013,Remy2014,De_Vis2019,Wiseman2017,Kahre2018}. The process of dust evolution consists of the dust production and destruction. The dust production is mainly motivated by dust formation in stellar winds and supernovae and grain growth in ISM. The dust destruction is mainly related to star formation depletion, thermal sputtering and supernova destruction. Besides, gas inflows and outflows also affect the amount of dust. Among these mechanisms, grain growth increases $\zeta$ while dust destruction have the opposite effect. If there are no grain growth and dust destruction, the $\zeta$ is a constant that depends on the yield of dust in stellar winds and supernovae. When there is a balance between dust grain growth and destruction, $\zeta$ is also a constant \citep[i.e.][]{De_Vis2017, Mattsson2012}. On the other hand, in dense and metal rich environments (e.g. in the inner regions of molecular clouds), the dust growth effect outweighs the destruction of dust particles, we would expect that $\zeta$ will increase with metallicity. In this paper, we propose that the stellar mass of host galaxy plays a more fundamental role in $\zeta$ of $\HII$ regions. In this Section, we first discuss the $\zeta\mbox{-}Z$ relation implied by $\rm{M_*}\mbox{-}\zeta$ relation in our study (Section \ref{sec:zeta-Z}) and then explore possible physical mechanisms behind the $\rm{M_*}\mbox{-}\zeta$ relation we propose (Section \ref{sec:zeta-M}). 

\subsection{\texorpdfstring{$\zeta\mbox{-}\rm{Z}$}{zeta-z} relation}
\label{sec:zeta-Z}

\begin{figure}

    \centering
    \gridline{\fig{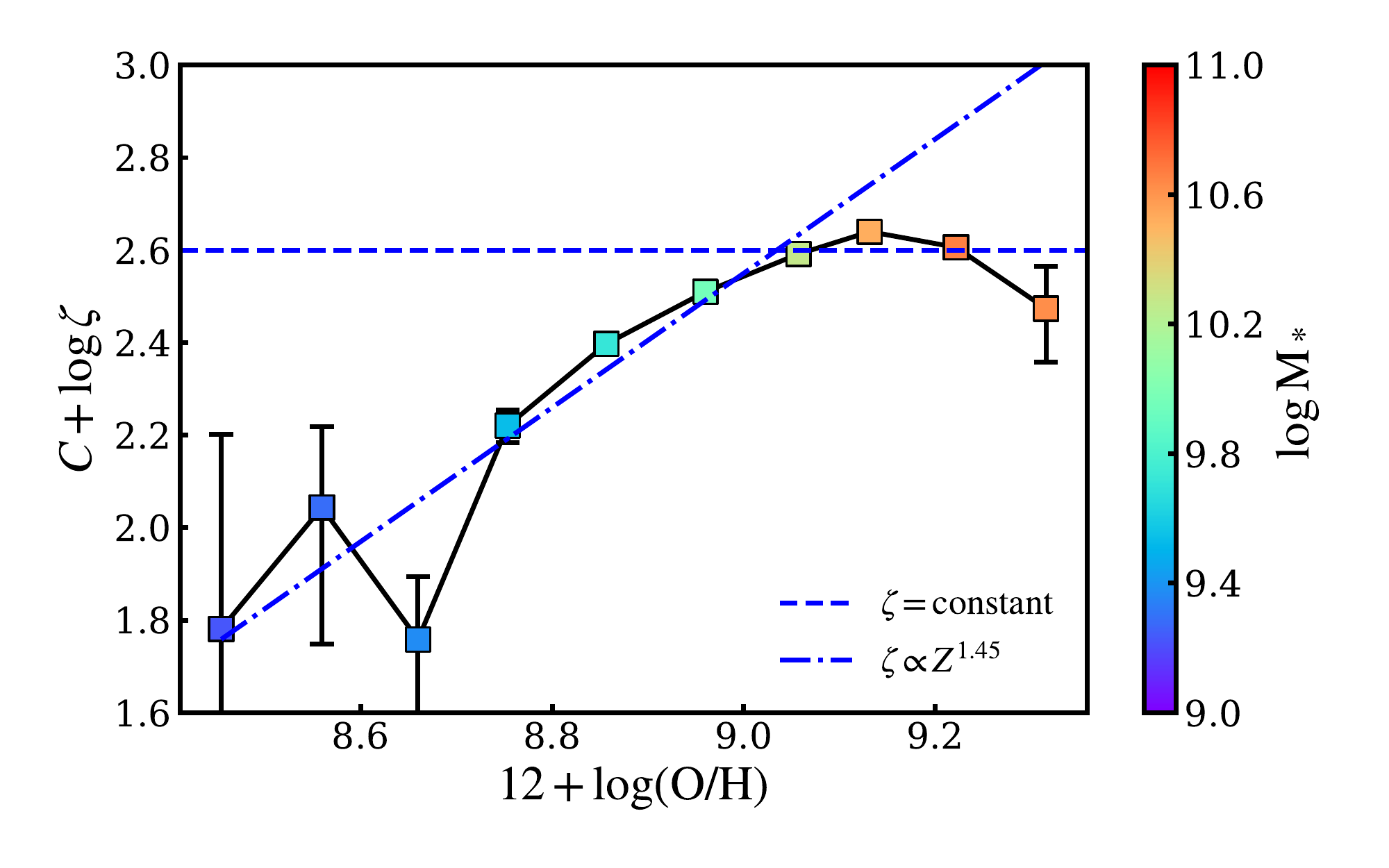}{3.5in}{}
          }
    \caption{ The dependence of $\log \zeta$ on 12+$\log (\rm{O/H})$ with stellar mass color-coded, where the $\log \zeta$ is interpreted by $\log \tau_{\rm{cl}}-\log(\rm{O/H})$. The dot-dashed line has a slope of 1.45 and the dashed line represent the DTM is a constant. }
    \label{fig4}
\end{figure}

Many studies have discussed the dependence of $\zeta$ on $Z$ \citep[e.g.][]{Silva1998,James2002,Clark2016,Yajima2015,Camps2015,Somerville2012,Ma2019,Katz2019}. To have a good comparison of our study with earlier results, we also explore the dependence of $\zeta$ on $Z$ only. In specific, we divide our sample SFGs into 10 metallicity bins in the range of $8.4<12+\log(\rm{O/H})<9.4$ with bin-width of 0.1 dex and then calculate the median $\tau_{\rm{cl}}$ of the sample galaxies in each metallicity bin. Folowing Equation \ref{tcl3}, we calculate $\zeta$ and then show its dependence on metallicity ($12+\log(\rm{O/H})$) in Figure \ref{fig4}, where the median $\log \rm{M_*}$ of sample galaxies in each metallicity bin is color coded. We see that $\zeta$ increases monotonically with $\log(\rm{O/H})$ at low metallicity range ($12+\log(\rm{O/H}) < 9.0$). At high metallicity end ($12+\log(\rm{O/H}) > 9.0$), $\zeta$ is roughly a constant. Such a trend has been reported in early studies \citep[e.g.][]{Remy2014}. 

For more quantitative comparisons, we over-plot two suggested $\zeta\mbox{-}Z$ relations in the literature in Figure \ref{fig4}. One is the relation $\zeta \sim \rm{constant}$ \citep{Issa1990,Lisenfeld1998} represented by the dashed line, where the intercept has been adjusted to fit the observed $\log \zeta$ at high metallicity. The other is the non-linear relation $\zeta \propto Z^{1.45}$  \citep[e.g.][]{De_Vis2019,LiQi2019}.  This relation is shown by the dot-dashed line and the intercept also has been adjusted to fit the observed $\zeta\mbox{-}Z$ relation at low metallicity. As can be seen, these two known relations agree well with our results at low and high metallicity parts respectively.

According to the dependence of  $\tau_{\rm{cl}}$ on the two dimensional $\rm{M_*}$ and $Z$ bins shown in the top-right panel of Figure \ref{fig3}, the slope of 1.45 between $\log \zeta$ and $\log(\rm{O/H})$ at the low metallicity end shown in Figure \ref{fig4} is more of the relation between $\tau_{\rm{cl}}$ and $\rm{M_*}$, which also can be clearly viewed from the color (stellar mass) of data points. At high metallicity end, $\rm{M_*}\mbox{-}Z$ relation becomes flat and $\rm{M_*}\mbox{-}\tau_{\rm{cl}}$ relation for high mass (metallicity) galaxies also becomes flat. As a result, $\zeta$ is a constant. In summary, in our study, the observed $\zeta\mbox{-}Z$ relation is a joint result of the  $\rm{M_*}\mbox{-}Z$ (Equation \ref{eq6}) and  $\rm{M_*}\mbox{-}\zeta$ (Equation \ref{DTMM}) relation.

\subsection{Physical implications of \texorpdfstring{$\rm{M_*}\mbox{-}\zeta$}{zeta-M} relation}
\label{sec:zeta-M}

The $\rm{M_*}\mbox{-}\zeta$ relation we have derived have important physical implications on the  assembly history of different components (e.g. stellar population, metal, dust) of galaxies. 

The $\rm{M_*}\mbox{-}Z$ relation of local SFGs could be a result of higher surface density of more massive disk galaxies \citep[][]{Chang2010,Belfiore2017}. Moreover, the star formation history of present-day galaxies shows a ``downsizing" behavior, where the stars in more massive galaxies tend to be formed earlier and over a shorter time-span \citep{Neistein2006}. Therefore, the higher $Z$ of more massive galaxies  could also be a result of the higher star formation efficiency at higher redshift, when the surface gas density of galaxies is averagely higher \citep{Fu2009}. In general, the mass assembly history of galaxies is in the order of a few Gyrs, which is much longer than the dust grain growth time-scale in molecular clouds ($\sim 10^7-10^8$ year \citep{Galliano2022}). Therefore, considering that more massive galaxies assemble their metals earlier and the shorter dust grain growth time-scale in higher metallicity regions, it is natural that more massive galaxies have a higher $\zeta$ today. 

On the other hand, the higher metallicity of more massive galaxies are possibly related to their deeper gravitational potential, where the outflows are suggested to be more difficult to escape from the host galaxy \citep{Chang2010, Chisholm2017}.  Considering heavier mass of metal particles, for both the momentum and energy driven outflows, the metal particles naturally have lower outflow velocity and outflow rate than that of gas \citep{Pandya2021}. Predictably, the outflow fraction of dust grains would be even smaller than that of metal particles, as the dust grains are much heavier than metal particles. Therefore, the outflow scenario also provides a reasonable explanation to the lower $Z$ and $\zeta$ in less massive galaxies, where the outflow is more prominent. 

At a given stellar mass, our study shows that  $\zeta$ is independent of gas-phase metallicity. This phenomena is possibly related to the physical condition and lifetime of $\HII$ regions. For the out layer of $\HII$ regions (or diffuse molecular cloud environments),  the time-scale of dust growth  ($\sim 100 \rm{Myr}$ \citep[][]{Galliano2022}) is much longer than the lifetime of $\HII$ regions ($\sim$ 10 $\rm{Myr}$), and the temperature ($\sim 10^4 K$) there is a large offset from the requirement of dust sputtering or sublimation($\sim 10^6 K$). 
Besides, the dust destruction effect by supernova can also be neglected since living OB stars are needed for $\HII$ regions. Therefore, $\zeta$ has not enough time to evolve during the short lifetime of $\HII$ regions. On the other hand, for galaxies at given stellar mass, the linear correlation between  metallicity and optical depth is probably related to the stochastic feeding (e.g. metal-poor gas accretion) of SFG, which is a transient process and has been used to explain the local anti-correlation between star formation rate(SFR) and gas-phase metallicity \citep{Sanchez2019}. In this scenario, the stochastic feeding of metal-poor gas in short time-scale does not change either the total amount of metal or dust, that is, this kind of inflow causes a simultaneous decrease of both metallicity and DTG ($\tau_{cl}$) and therefore does not affect $\zeta$.

\section{summary}
\label{sec:conclusion}

In this study, by applying a simplified version of the Chocolate Chip Cookie model of \citet{Lu2022} on a sample of face-on star-forming galaxies selected from SDSS, we obtain the typical dust optical depth $\tau_{cl}$ of the $\HII$ regions for galaxies in different stellar mass and gas-phase metallicity bins. By investigating the dependence of $\tau_{cl}$ on $\rm{M_*}$ and $Z$ and linking $\tau_{cl}$ to DTG of galaxies, we generate the following conclusions on DTG and DTM of star-forming galaxies.

We find that $\tau_{\rm{cl}}$ increases with $\rm{M_*}$ faster than $Z$ so that DTM ($\zeta \sim \tau_{\rm{cl}}/Z$) increases with $\rm{M_*}$ of galaxies. At a given stellar mass, the residual $\Delta \log \tau_{\rm{cl}}$ is linearly correlated with $\Delta \log(\rm{O/H})$, implying a constant $\zeta$. Our results show that the stellar mass of galaxies is the first parameter in the DTM-$Z$ relation of galaxies, like in many other scaling relations\citep[e.g.][]{Shen2003,Kauffmann2003b,Peng2010}. The  $\zeta\mbox{-}Z$ relation discussed in literature is a joint result of the $\rm{M_*}\mbox{-}\zeta$ and $\rm{M_*}\mbox{-}Z$ relation.

\section*{acknowledgments}

We thank the anonymous referee for the helpful and constructive comments that improve the paper. This work is supported by  the National Natural Science Foundation of China (No. 12073059 $\&$ No. U2031139) , the National Key R$\&$D Program of China (No. 2019YFA0405501), an the Program of Shanghai Academic/Technology Research Leader (22XD1404200). We also acknowledge the science research grants from the China Manned Space Project with NO. CMS-CSST-2021-A04, CMS-CSST-2021-A07, CMS-CSST-2021-A08, CMS-CSST-2021-A09, CMS-CSST-2021-B04. F.T.Y. acknowledges support by the Funds for Key Programs of Shanghai Astronomical Observatory (No. E195121009) and the Natural Science Foundation of Shanghai (Project Number: 21ZR1474300). 

Funding for the SDSS and SDSS-II has been provided by the Alfred P. Sloan Foundation, the Participating Institutions, the National Science Foundation, the U.S. Department of Energy, the National Aeronautics and Space Administration, the Japanese Monbukagakusho, the Max Planck Society, and the Higher Education Funding Council for England. The SDSS Web Site is http://www.sdss.org/.

The SDSS is managed by the Astrophysical Research Consortium for the Participating Institutions. The Participating Institutions are the American Museum of Natural History, Astrophysical Institute Potsdam, University of Basel, University of Cambridge, Case Western Reserve University, University of Chicago, Drexel University, Fermilab, the Institute for Advanced Study, the Japan Participation Group, Johns Hopkins University, the Joint Institute for Nuclear Astrophysics, the Kavli Institute for Particle Astrophysics and Cosmology, the Korean Scientist Group, the Chinese Academy of Sciences (LAMOST), Los Alamos National Laboratory, the Max-Planck-Institute for Astronomy (MPIA), the Max-Planck-Institute for Astrophysics (MPA), New Mexico State University, Ohio State University, University of Pittsburgh, University of Portsmouth, Princeton University, the United States Naval Observatory, and the University of Washington.

\bibliography{ref}
\bibliographystyle{aasjournal}

\end{CJK*}
\end{document}